\documentclass[twocolumn]{aastex631}
\usepackage{graphicx,times}            
\usepackage{natbib}
\usepackage{amssymb,amsmath}
\usepackage[figuresright]{rotating}
\usepackage{rotating} 
\usepackage{color}
\usepackage{multirow}
\usepackage{longtable,booktabs}
\usepackage[titletoc]{appendix}
\usepackage{rotfloat}
\usepackage{threeparttable}
\usepackage{times}
\usepackage{verbatim}
\usepackage{subfigure}
\usepackage{longtable}
\usepackage{ulem}
\usepackage{soul}

\shorttitle{Discovery of an Extremely $r$-process-enhanced Thin-disk Star}
\shortauthors{X. J. Xie et al.}

\graphicspath{{./}{figures/}}

\begin{document}

\title{Discovery of an Extremely $r$-process-enhanced Thin-disk Star with [Eu/H] $= +0.78$}

\author[0000-0002-4440-4803]{Xiao-Jin Xie}
\affiliation{CAS Key Laboratory of Optical Astronomy, National Astronomical Observatories, Chinese Academy of Science, Beijing 100101, China}
\affiliation{School of Astronomy and Space Science, University of Chinese Academy of Sciences, Beijing 100049, China}

\author[0000-0002-0349-7839]{Jianrong Shi}\thanks{corresponding author: sjr@bao.ac.cn}
\affiliation{CAS Key Laboratory of Optical Astronomy, National Astronomical Observatories, Chinese Academy of Science, Beijing 100101, China}
\affiliation{School of Astronomy and Space Science, University of Chinese Academy of Sciences, Beijing 100049, China}
\affiliation{School of Physics and Technology, Nantong University, Nantong 226019, China}

\author[0000-0002-8609-3599]{Hong-Liang Yan}
\affiliation{CAS Key Laboratory of Optical Astronomy, National Astronomical Observatories, Chinese Academy of Science, Beijing 100101, China}
\affiliation{School of Astronomy and Space Science, University of Chinese Academy of Sciences, Beijing 100049, China}
\affiliation{Institute for Frontiers in Astronomy and Astrophysics, Beijing Normal University, Beijing 102206, China}

\author[0000-0002-6448-8995]{Tian-Yi Chen}
\affiliation{CAS Key Laboratory of Optical Astronomy, National Astronomical Observatories, Chinese Academy of Science, Beijing 100101, China}
\affiliation{School of Astronomy and Space Science, University of Chinese Academy of Sciences, Beijing 100049, China}

\author[0000-0002-0084-572X]{Carlos Allende Prieto}
\affiliation{Instituto de Astrofísica de Canarias, Vía Láctea S/N, E-38205 La Laguna, Tenerife, Spain}
\affiliation{Universidad de La Laguna, Departamento de Astrofísica, E-38206 La Laguna, Tenerife, Spain }

\author[0000-0003-4573-6233]{Timothy C. Beers}
\affiliation{Department of Physics and JINA Center for the Evolution of the Elements, University of Notre Dame, Notre Dame, IN 46556, USA}

\author[0000-0001-5193-1727]{Shuai Liu}
\affiliation{CAS Key Laboratory of Optical Astronomy, National Astronomical Observatories, Chinese Academy of Science, Beijing 100101, China}

\author[0000-0002-6647-3957]{Chun-Qian Li}
\affiliation{CAS Key Laboratory of Optical Astronomy, National Astronomical Observatories, Chinese Academy of Science, Beijing 100101, China}
\affiliation{School of Astronomy and Space Science, University of Chinese Academy of Sciences, Beijing 100049, China}

\author[0000-0001-6898-7620]{Ming-Yi Ding}
\affiliation{CAS Key Laboratory of Optical Astronomy, National Astronomical Observatories, Chinese Academy of Science, Beijing 100101, China}
\affiliation{School of Astronomy and Space Science, University of Chinese Academy of Sciences, Beijing 100049, China}

\author[0009-0006-3386-4632]{Yao-Jia Tang}
\affiliation{CAS Key Laboratory of Optical Astronomy, National Astronomical Observatories, Chinese Academy of Science, Beijing 100101, China}
\affiliation{School of Astronomy and Space Science, University of Chinese Academy of Sciences, Beijing 100049, China}

\author[0009-0008-1319-1084]{Ruizhi Zhang}
\affiliation{CAS Key Laboratory of Optical Astronomy, National Astronomical Observatories, Chinese Academy of Science, Beijing 100101, China}
\affiliation{School of Astronomy and Space Science, University of Chinese Academy of Sciences, Beijing 100049, China}

\author[0009-0002-4282-668X]{Renjing Xie}
\affiliation{CAS Key Laboratory of Optical Astronomy, National Astronomical Observatories, Chinese Academy of Science, Beijing 100101, China}
\affiliation{School of Astronomy and Space Science, University of Chinese Academy of Sciences, Beijing 100049, China}

\accepted{July 6, 2024}
\submitjournal{ApJL}

\begin{abstract}

Highly $r$-process-enhanced stars are rare and usually metal-poor ([Fe/H] $ \textless $ $ - $1.0), and mainly populate the Milky Way halo and dwarf galaxies. This study presents the discovery of a relatively bright ($V = 12.72$), highly $r$-process-enhanced ($r$-II) star ([Eu/Fe] = $+$1.32, [Ba/Eu] = $ - $0.95), LAMOST~J020632.21$ + $494127.9. This star was selected from the Large Sky Area Multi-Object Fiber Spectroscopic Telescope (LAMOST) medium-resolution ($R \sim 7500$) spectroscopic survey; follow-up high-resolution ($R \sim 25,000$) observations were conducted with the High Optical Resolution Spectrograph (HORuS) installed on the Gran Telescopio Canarias (GTC). The stellar parameters (${T_{\rm eff}}$ = 4130\,K, $\rm log\,g $ = 1.52, $ \rm[Fe/H] $ = $ - $0.54, $\xi$ = 1.80 $ \rm{km\,{s^{-1}}} $) have been inferred taking into account non-local thermodynamic equilibrium (NLTE) effects. The abundances of [Ce/Fe], [Pr/Fe], and [Nd/Fe] are $+$0.19, $+$0.65 and $+$0.64, respectively, relatively low compared to the Solar $r$-process pattern normalized to Eu. 
This star has a high metallicity ([Fe/H] = $ - $0.54) compared to most other highly $r$-process-enhanced stars, and has the highest measured abundance ratio of Eu to H ([Eu/H] $= +0.78$). It is classified as a thin-disk star based on its kinematics, and does not appear to belong to any known stream or dwarf galaxy. 

\end{abstract}


\section{Introduction} \label{sec:intro}

The rapid neutron-capture process ($r$-process) is one of the fundamental nucleosynthetic pathways for producing the heaviest elements. Several astrophysical sites have been proposed to satisfy the physical conditions required for the $r$-process, such as core-collapse supernovae \citep[e.g.,][]{1974PThPh..51..726S, 1994AA...286..841W, 2010ApJ...712.1359F, 2015PhRvD..92j5020M, 2021ApJ...920L..32T},  magneto-rotational jet-driven supernovae  \citep[e.g.,][]{1985ApJ...291L..11S, 2008ApJ...680.1350F, 2015ApJ...810..109N, 2018JPhG...45h4001O, 2023MNRAS.518.1557R}, common envelop jet supernovae \citep[e.g.,][]{2019ApJ...878...24G,2023arXiv231008907J}, collapsars \citep{2019Natur.569..241S}, and neutron star mergers \citep[NSMs,][]{1974ApJ...192L.145L, 2014MNRAS.439..744R, 2017ARNPS..67..253T}. The detection of GW170817 \citep{2017PhRvL.119p1101A} and its electromagnetic counterpart shows that NSMs can indeed produce $r$-process elements \citep{2017Sci...358.1570D,2017Natur.551...80K, 2017Natur.551...67P, 2019Natur.574..497W}. Research is still ongoing to determine whether NSMs are the sole source of $r$-process elements; some studies suggest that the observed $r$-process-element abundance patterns requires multiple sites \citep[e.g.,][]{2018ApJ...855...99C, 2019ApJ...875..106C, 2018IJMPD..2742005H, 2019EPJA...55..203S, 2021ApJ...915...81B, 2021arXiv210703486F, 2021ApJ...920L..32T, 2021arXiv210205891Y, 2022ApJ...926L..36N, 2023MNRAS.525.2040E, 2023ApJ...943L..12K}. 

Generally, $r$-process-enhanced (RPE) stars are metal-poor, with [Fe/H] $<-$1.0 \citep[e.g.,][]{1995AJ....109.2757M, 2008ARA&A..46..241S, 2020ApJS..249...30H}, and they are mainly restricted to the Milky Way (MW) halo and dwarf galaxies \citep[e.g.,][]{2015ApJ...814...41H, 2016Natur.531..610J, 2021MNRAS.506.1850J, 2021A&A...650A.110M}. These stars are classified as $r$-I, $r$-II, or limited-$r$, depending on the level of $r$-process enrichment \citep{2005ARAA..43..531B, 2018ARNPS..68..237F, 2020ApJS..249...30H}. The dominant $r$-process patterns (Ba to Hf) of these stars are nearly identical (sometimes referred to as ``universaility"), although some star-to-star differences between the lightest and heaviest elements remain \citep[e.g.,][]{1994ApJ...431L..27S, 2017ApJ...837....8A, 2018ApJ...868..110S, 2022ApJ...936...84R}. Information on the elemental abundances and kinematics of these stars constrain their early enrichment history, which is crucial to understanding the origin of the $r$-process.

This work reports on the discovery of an RPE star  with [Fe/H] $>-$0.6 that is kinematically associated with the thin disk of the MW. Section \ref{sec:obs} introduces the observations, and Section \ref{sec:para} describes the determination of stellar atmospheric parameters. A detailed abundance analysis and estimates of uncertainties is presented in Section \ref{sec:abun}. 
A discussion of the implications of the existence of this star, and an old (but uncertain) estimate of its nucleosynthetic age based on the observed [Th/Eu] ratio, is provided in Section \ref{sec:disc}, followed by a summary of our conclusions in Section \ref{sec:conc}.

\section{Observations}  \label{sec:obs}  

The RPE candidate LAMOST J020632.21$ + $494127.9 (hereafter J\,0206$ + $4941)  was selected from the Large Sky Area Multi-Object Fiber Spectroscopic Telescope \citep[LAMOST,][]{2012RAA....12.1197C,2012RAA....12..723Z,2020arXiv200507210L,YAN2022100224} medium-resolution ($R \sim$ 7500) spectroscopic survey by matching the observed spectra to synthetic templates of the region around the \ion{Eu}{2} line at 6645.1\,\AA\ \citep{2021RAA....21...36C}. This star is relatively bright ($V<$ 13) and substantially more metal-rich compared to previously observed $r$-II stars.

The high-resolution spectrum was taken with the High Optical Resolution Spectrograph (HORuS) installed on the Gran Telescopio Canarias \citep[GTC,][]{2020MNRAS.498.4222T,2021NatAs...5..105A}.
The spectral wavelength coverage is from 3800 to 6900\,\AA\, with a resolving power of $R \sim 25,000$. The multi-order echelle spectra were extracted and wavelength calibrated based on Th-Ar hollow-cathode lamp exposures, using the HORuS automated pipeline the {\tt chain}\footnote{The {\tt chain} is written in the Interactive Data Language (IDL), runs in GDL, and is available from github.com/callendeprieto/chain.}. The signal-to-noise ratio (S/N) at 5600\,{\AA} is estimated to be 119, using the code \texttt{DER\_SNR} \citep{2008ASPC..394..505S}. Table~\ref{table1} lists the basic information of this target, including the details of the observations, kinematics, and stellar atmospheric parameters, as described in detail below.

\begin{table}[]
	\centering
		\caption{Target Information}\label{table1}
		\setlength{\tabcolsep}{3mm}
		\begin{threeparttable}
		\begin{tabular}{lc}
        \hline\noalign{\smallskip}
       Parameter   &    Value   \\     
       \hline\noalign{\smallskip}
       Name  &   LAMOST J\,020632.21$ + $494127.9 \\
       R.A. (J2000)        &     02:06:32.211    \\
       Decl. (J2000)        &     +49:41:27.914    \\
       $V$ mag &     12.72 $\pm$ 0.24                        \\
       obsDate $_{\rm LAMOST}$  & 2018-01-01   \\
       obsDate $_{\rm GTC}$ & 2019-11-18   \\
       $\rm{{RV}_{helio}}$ (km s$^ {-1} $) &   $ - $30.86 $\pm$ 0.67   \\
       $\rm{ Parallax} $ (mas)    &      0.3106 $\pm$ 0.0156  \tnote{a}        \\
       $\rm{{U}_{LSR}}$ (km s$^ {-1} $)  &    31.11 $\pm$ 0.49      \\
       $\rm{{V}_{LSR}}$ (km s$^ {-1} $)  &    $ - $11.89 $\pm$ 0.50      \\
       $\rm{{W}_{LSR}}$ (km s$^ {-1} $)  &    9.65 $\pm$ 0.35      \\ 
       ${{r}_{apo}}$ (kpc)            &    10.79 $\pm$ 0.17      \\
       ${{r}_{peri}}$ (kpc)           &    9.68 $\pm$ 0.06      \\
       ${{Z}_{max}}$ (kpc)            &    0.66 $\pm$0.03      \\
       $ecc$                              &    0.054$\pm$0.004        \\
       $ E  $ ($ 10^3 $ km$^ {2} $ s$^ {-2} $) & $ - $144.31 $\pm$ 0.60  \\
       $ J_r $ (kpc km s$^ {-1} $)   &  4.82 $\pm$ 0.85 \\
       $ J_\phi $ (kpc km s$^ {-1} $)   & 2353.05 $\pm$ 24.66  \\
       $ J_z $ (kpc km s$^ {-1} $)   & 8.27 $\pm$ 0.51  \\
      $ {T_{\rm eff}} $ (K)         &    4078 $\pm$ 100   \tnote{b}         \\
      $\rm log\,g  $ (cgs)            &     \,1.62 $\pm$ 0.08\tnote{b}           \\     
       $ {T_{\rm eff}} $ (K)             &     4130 $\pm$ 100 \tnote{c}          \\      
       $\rm log\,g  $ (cgs)                  &    1.52 $\pm$ 0.20    \tnote{c}         \\
       $ \rm[Fe/H] $                  &    $ - $0.54 $\pm$ 0.10  \tnote{c}          \\
       $\xi_{\rm t}\rm (km\,{s^{-1}})$         &    1.80$ \pm$ 0.20	  \tnote{c}        \\
        \noalign{\smallskip}\hline	
		\end{tabular}
    \begin{tablenotes}
    	\footnotesize
    	\item[a] \cite{2023AA...674A...1G}
    	\item[b] Determined by a photometric method. 
    	\item[c] Determined by a spectroscopic approach.
    \end{tablenotes}
\end{threeparttable}
\end{table}

\begin{figure*}
	\centering
	\includegraphics[width=0.95\linewidth]{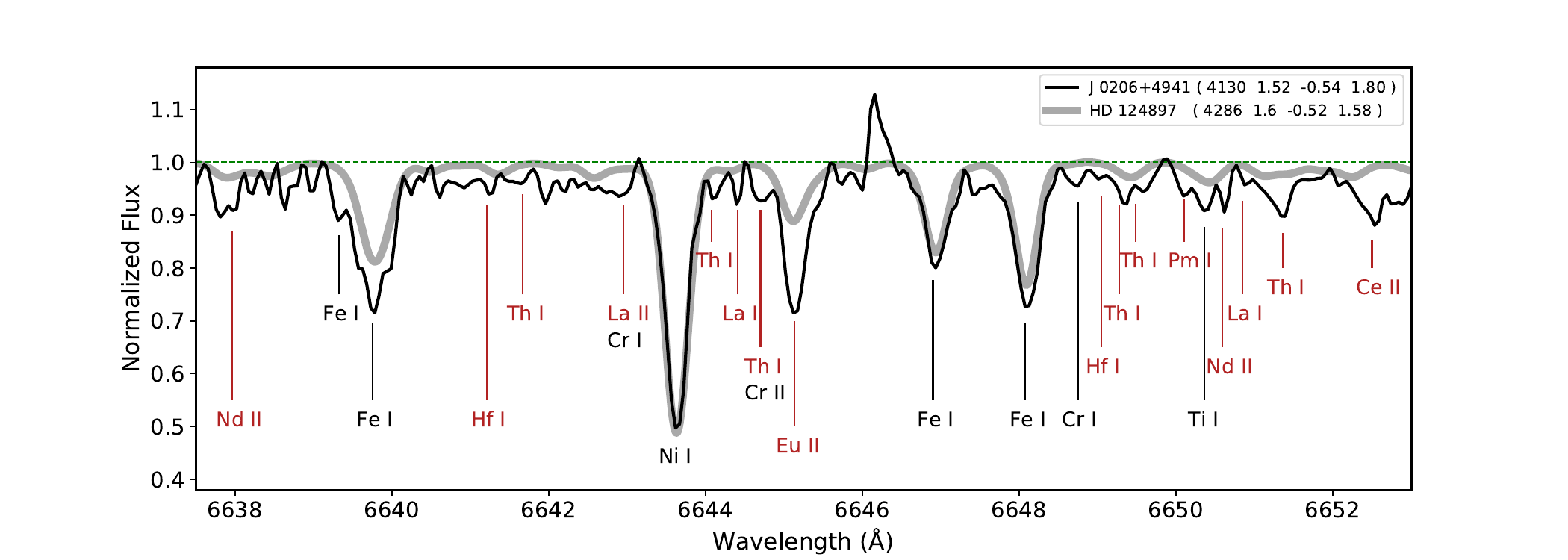}
	\vspace*{-2mm}
	\caption{Comparison of high-resolution spectra of J\,0206$ + $4941 and a moderately 
 $r$-process-enhanced ($r$-I) star, HD~124897 ([Eu/Fe] = +0.45, \citealt{2019A&A...631A.113F}) with similar stellar parameters. The lines of heavy elements are labeled in red, and lines of lighter elements are labeled in black. The stellar parameters ( ${T_{\rm eff}}$, \,log g, \,[Fe/H], \,$\xi$ ) are shown in the legend.} 
	\label{comparison}
\end{figure*}

The high-resolution spectrum of J\,0206$ + $4941 around 6645\,{\AA} is shown in Figure \ref{comparison}, along with the spectrum of HD~124897, a star with similar atmospheric parameters (${T_{\rm eff}}$ = 4286 K, log g = 1.6 cgs, [Fe/H] = $-$0.52, and $\xi$ = 1.58 km $\rm{s^{-1}}$, \citealt{2014AA...566A..98B}). The slightly higher ${T_{\rm eff}}$ of HD~124897 leads to stronger Fe-group lines. The lines of heavy elements are labeled in red. HD~124897 has a moderate 
$r$-process enhancement of [Eu/Fe] = +0.45 \citep{2019A&A...631A.113F}, and is classified as an $r$-I star. As shown in the figure, the Eu~II line of our candidate star is significantly stronger than that of HD~124897. Many lines of heavy elements are weak or absent in the spectra of HD~124897, but they are strong in the spectra of our candidate star, indicating that J\,0206$ + $4941 has significant 
$r$-process enhancement.

\section{Stellar Atmospheric Parameters}\label{sec:para}

 We first derive initial estimates of effective temperature and surface gravity of J\,0206+4941 using a photometric approach. For comparison, we also determine the atmospheric parameters using a spectroscopic approach: effective temperature ($T_{\rm eff}$), surface gravity (log\,$\emph{g}$), metallicity ([Fe/H]), and microturbulent velocity ($\xi_{\rm t}$).

\subsection{Photometric Approach}

We estimate the effective temperature of our target star based on the photometric color index ($V - K$)$_0$ with the empirical calibration relations given by \citet{1999AAS..140..261A}. An adopted reddening $E(B - V)$ of $0.1518\,\pm\,0.0083$ is derived using the \texttt{GALEXTIN} online tool \citep{2021MNRAS.508.1788A} with the \citet{2019ApJ...887...93G} 3D dust map. The $E(V - K)$ reddening is determined using the conversions from \citet{2005ApJ...626..465R}. 
The surface gravity is calculated using the online code \texttt{PARAM 1.3} \citep{2006AA...458..609D}, based on the relationships between bolometric flux, temperature, mass, and log\,$\emph{g}$. The calculation requires estimates of $T_{\rm eff}$, [Fe/H], $V$ magnitude, and parallax, where we use [Fe/H] measured by the spectroscopic approach described below, the $V$ magnitude from \citet{2013AJ....145...44Z}, and the parallax from \textit{Gaia} DR3 \citep{2023AA...674A...1G}.

\subsection{Spectroscopic Approach}

The line list and atomic data of \ion{Fe}{1} and \ion{Fe}{2} lines are taken from \cite{2018NatAs...2..790Y}. Only lines with equivalent widths (EWs) between 20\,m\AA\, and 120\,m\AA,  and not heavily blended with other species, have been used. This selection results in 45 \ion{Fe}{1} and 7 \ion{Fe}{2} lines, respectively. The abundances and EWs of individual Fe lines are analyzed by the interactive IDL code Spectrum Investigation Utility \citep[\texttt{SIU},][]{1999PhDT.......216R} using the MARCS stellar atmospheric models \citep{2008AA...486..951G}. The effects of non-local thermodynamic equilibrium (NLTE) are considered specifically for each \ion{Fe}{1} line, based on \citet{2011A&A...528A..87M}, following the method described in \citet{2015ApJ...808..148S}. 

A spectroscopic estimate of $T_{\rm eff}$ is obtained from the excitation equilibrium of \ion{Fe}{1} lines with excitation energies (E$_{\rm exc}$) higher than 2.0\,eV \citep{2015ApJ...808..148S}, log~$g$ is estimated from the ionization balance of \ion{Fe}{1} and \ion{Fe}{2}, and $\xi_{\rm t}$ is obtained by forcing the iron abundances from \ion{Fe}{1} lines to be independent of their EWs. We adjust $ T_{\rm eff} $, log~$g$, and $\xi_{\rm t}$ by restricting the slope of log$\epsilon$(\ion{Fe}{1}) versus E$_{\rm exc}$ lower than 0.001\,dex\,eV$^{-1}$, the difference between [\ion{Fe}{2}/H] and [\ion{Fe}{1}/H] within 0.1\,dex, and the slope of log$\epsilon$(\ion{Fe}{1}) versus log(EW) less than 0.01. 

The results determined by both methods are listed in Table~\ref{table1}.
The difference between the photometric and spectroscopic effective temperatures is only 52\,K, and the difference in surface gravities between the two methods is 0.10\,dex. We adopt the spectroscopic parameters in the following analysis.

\begin{figure*}[]
	\centering
	\gridline{\fig{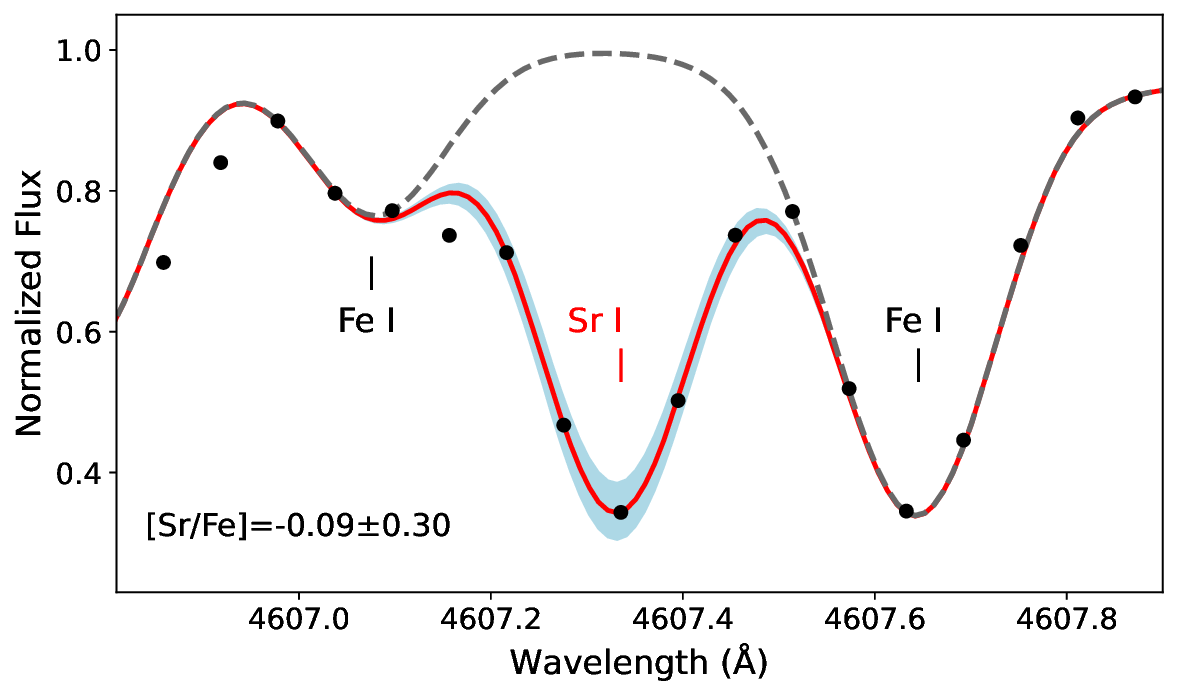}{0.48\textwidth}{}
		\fig{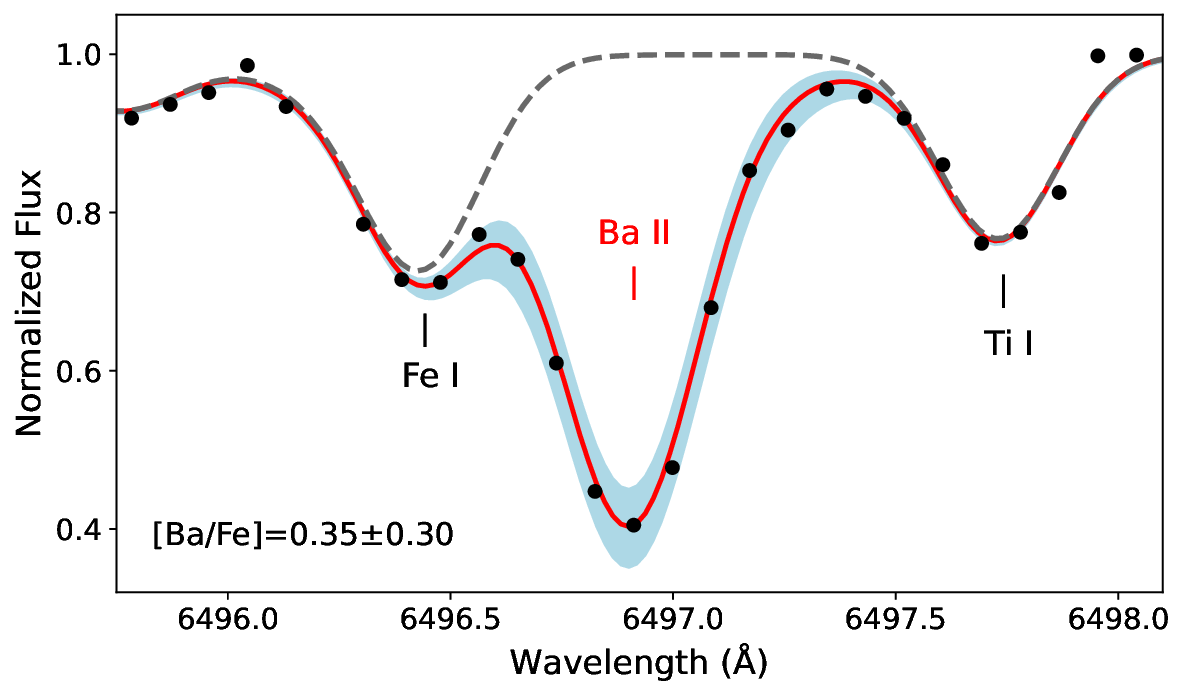}{0.48\textwidth}{}
	}
	\vspace*{-9mm}
	\gridline{	\fig{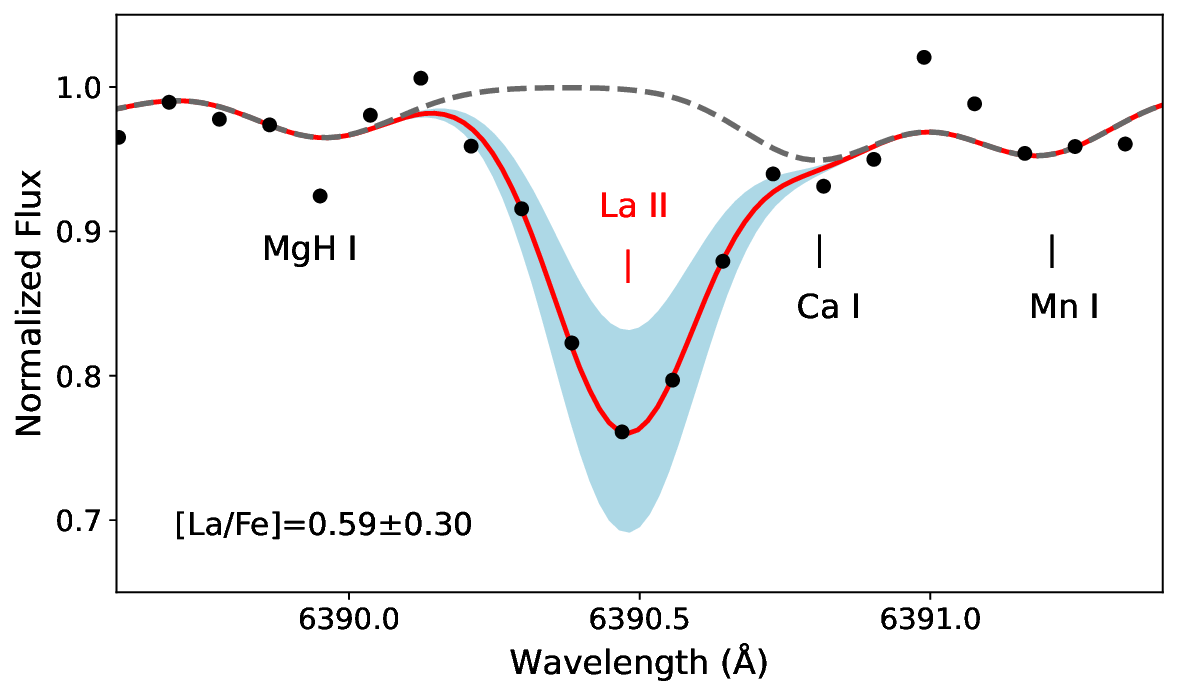}{0.48\textwidth}{}
		\fig{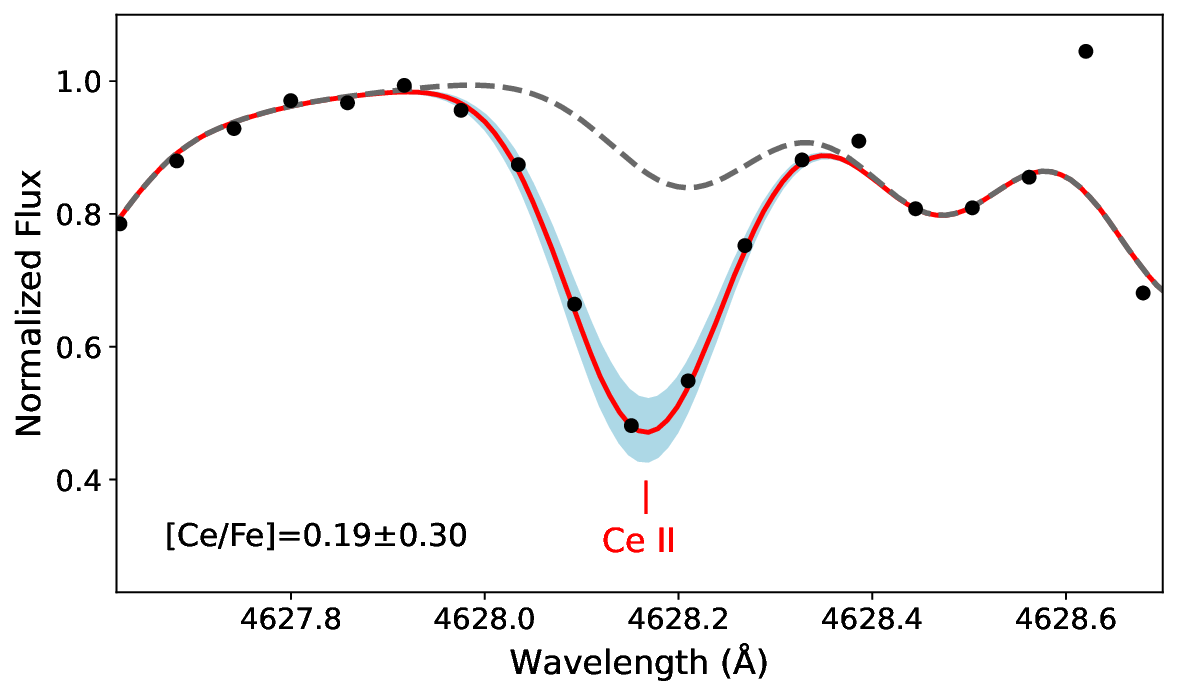}{0.48\textwidth}{}
	}
 \vspace*{-9mm}
	\gridline{	\fig{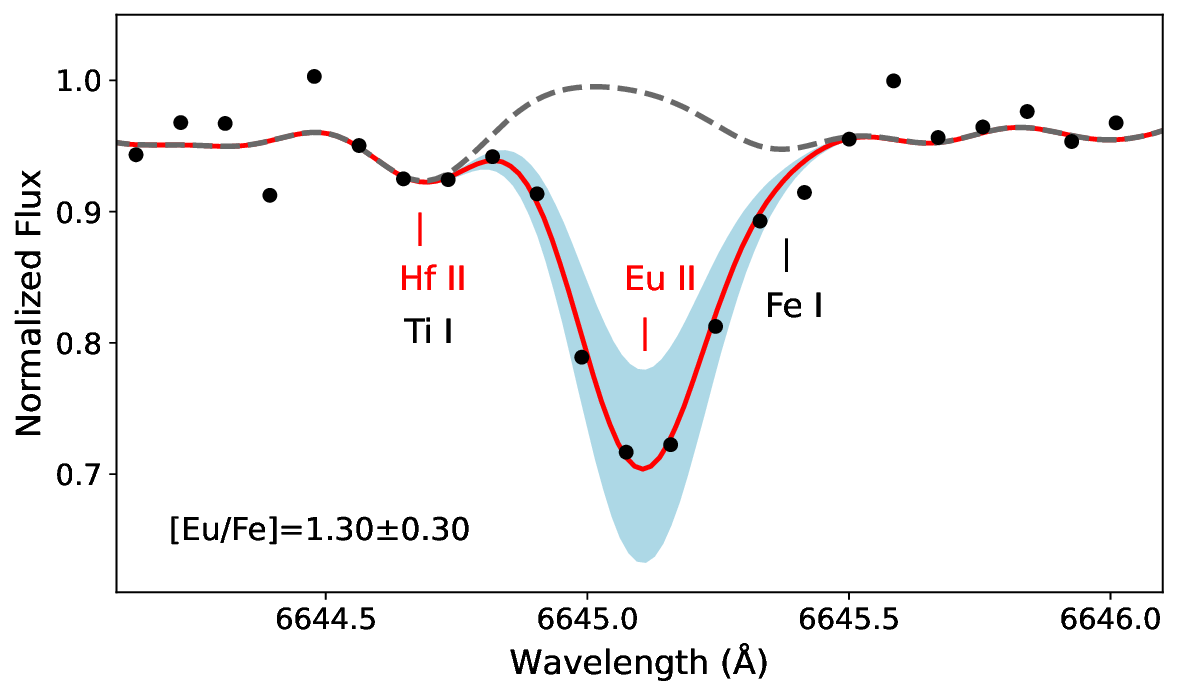}{0.48\textwidth}{}
		\fig{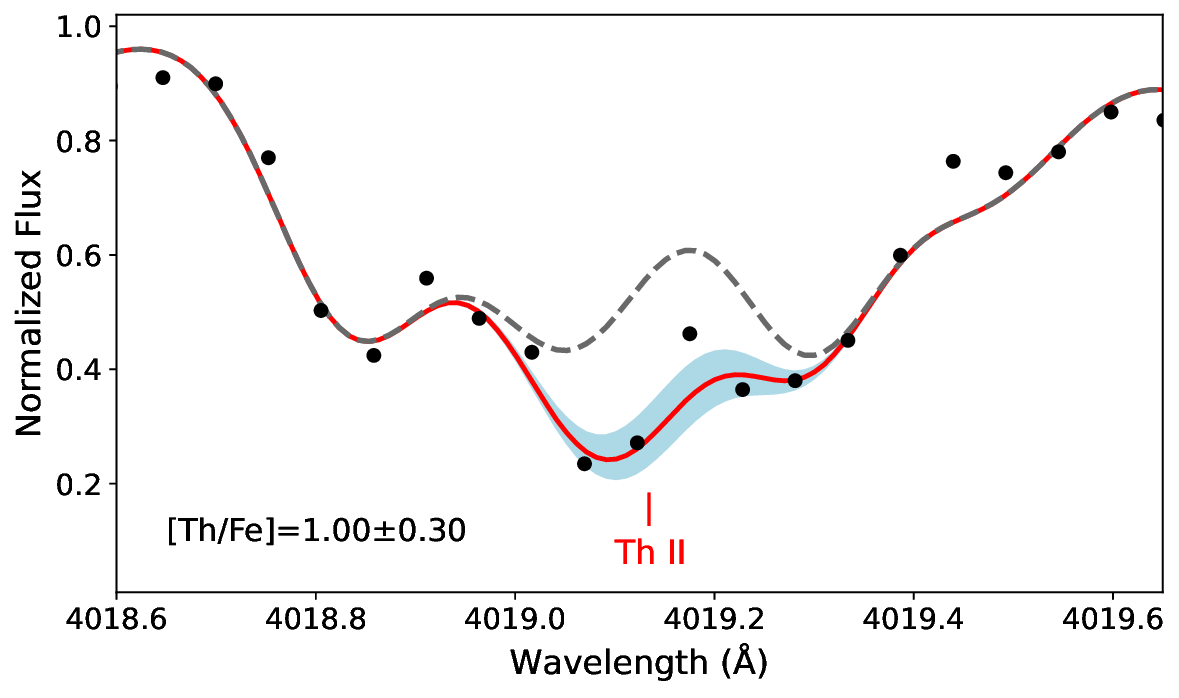}{0.48\textwidth}{}
	}
	\vspace*{-8mm}
	\caption{Spectral synthesis for the abundance determinations of Sr, Ba, La, Ce, Eu and Th. The red line indicates the best-fit synthesis for the line(s) of interest. The blue shaded regions indicate $\pm$0.30 dex deviations from the best fit. The gray-dashed lines indicate synthetic spectra in the absence of the element of interest. } 
	\label{syn}
\end{figure*}

\section{Analysis of Elemental Abundances} \label{sec:abun}

\subsection{Abundance Estimates}

We use a spectrum-synthesis approach for the abundance determination. The line list and atomic data are taken from \citet{2016ApJ...833..225Z} and \citet{2018ApJ...865..129R}; we have slightly revised the log\,$gf$ values by fitting the Solar spectrum \citep{1984Solar}. The influence of hyperfine structure (HFS) is considered for Sc \citep{2008A&A...481..489Z}, Cu \citep{2014ApJ...782...80S}, Sr \citep{1983HyInt..15..177B}, Ba \citep{1995all..book.....K}, and Eu \citep{2001ApJ...563.1075L}. We calculate the theoretical line profiles by \texttt{SIU}, using the MARCS stellar model atmospheres \citep{2008AA...486..951G}. The Solar abundances are adopted from \citet{2009ARAA..47..481A}. 
 
The carbon abundance is estimated from the molecular CH $G$-band region around 4310\,\AA; the line list we adopted comes from \citet{2014AA...571A..47M}. Among the light odd-$Z$ elements, we can obtain abundance estimates of Na and Al. Four \ion{Na}{1} lines at 5682\,{\AA}, 5688\,{\AA}, 6154\,{\AA}, and 6160\,{\AA}, and two \ion{Al}{1} lines at 6696\,{\AA} and 6698\,{\AA} have been used.
The $\alpha$-elements with detectable lines include Mg, Si, Ca, and Ti, while the [\ion{O}{1}] line at 6300\,{\AA} is too weak to derive a reliable abundance. NLTE effects have been taken into consideration, based on \citet[Na]{2004A&A...413.1045G}, \citet[Al]{1996A&A...307..961B}, \citet[Mg]{2013A&A...550A..28M}, \citet[Si]{2008A&A...486..303S}, \citet[Ca]{2007A&A...461..261M}, \citet[Ti]{2016MNRAS.461.1000S}, and \citet[Cu]{2014ApJ...782...80S}. In addition, the NLTE corrections for lines of Cr~I, Mn~I, and Co~I are obtained from an online spectral tool \footnote{https://nlte.mpia.de \citep{2019NLTE}}.
 
The neutron-capture elements frequently have their absorption lines blended with other elements. This is especially the case for relatively metal-rich stars, as strong Fe lines often lead to strong blends.  Figure \ref{syn} shows a sample of spectral synthesis for the \ion{Sr}{1}, \ion{Ba}{2}, \ion{La}{2}, \ion{Ce}{2}, \ion{Eu}{2}, and \ion{Th}{2} lines.  We use the \ion{Sr}{1} line at 4607\,{\AA} to determine the Sr abundance. There are numerous very strong absorption lines for elements, such as Fe and Ti lines around 4077\,{\AA} and 4161\,{\AA}, making it difficult to distinguish the absorption lines of \ion{Sr}{2}. The abundance of Ba is derived from the \ion{Ba}{2} lines at 4554\,{\AA}, 4934\,{\AA}, 6141\,{\AA}, and 6496\,{\AA}. The NLTE effects have been taken into account \citep{1999A&A...343..519M}. The abundance of Eu is derived from the \ion{Eu}{2} lines at 4129\,{\AA}, 4205\,{\AA}, 6049\,{\AA}, and 6645\,{\AA}, with the NLTE effects considered \citep{2000A&A...364..249M}. As the lines at 4129\,{\AA} and 4205\,{\AA} are heavily blended, only upper limits on Eu are obtained from these two lines. The abundance of Ce is derived from the \ion{Ce}{2} lines at 4418\,{\AA}, 4628\,{\AA}, 5247\,{\AA}. The abundance of Th is estimated by using the \ion{Th}{2} line at 4019\,{\AA}. The abundances for a total of 15 neutron-capture elements, including Sr, Y, Zr, Ba, La, Ce, Pr, Nd, Sm, Eu, Gd, Dy, Lu, Hf, and Th have been determined; the results are listed in Table~\ref{table2}.  
 
\subsection{Uncertainties} 

To estimate the uncertainty of the abundance determinations, two types of error sources are considered: systematic errors (${\sigma_{sys}}$, due to errors in the stellar atmospheric parameters), and statistical errors (${\sigma_{stat}}$, due to the line measurements themselves).  
Table~\ref{table2} lists the abundance differences when changing $T_{\rm eff}$ by $+ 100$\,K, log~$g$ by $+ 0.2$ dex, [Fe/H] by $+ 0.1$ dex, and $\xi_{\rm t}$ by $+ 0.2$ km $\rm{s^{-1}}$. The ${\sigma_{sys}}$ is calculated by taking the square root of the quadratic sum of the errors associated with the atmospheric parameters. The ${\sigma_{stat}}$ is the line-to-line dispersion divided by $\sqrt{N}$, where $N$ is the number of lines used for a given element.  Finally, we computed the total error, represented as ${\sigma_{Total}}$, by summing ${\sigma_{sys}}$ and ${\sigma_{stat}}$ in quadrature.
  
  \begin{table*}
	\centering
	\caption[]{Elemental Abundances and Uncertainties\label{table2}}
	\setlength{\tabcolsep}{2mm}
	\begin{tabular}{ccrrrrrrccr}
		\hline\noalign{\smallskip}
		\multirow{2}*{Element} & \multirow{2}*{N} & [X/Fe] &  [X/Fe] & {${\sigma_{stat}}$}  &$\Delta\,{T_{\rm eff}}$  &$\Delta$\,log g   &$\Delta$\,[Fe/H]   &$\Delta\,\xi$     & ${\sigma_{sys}}$  &  ${\sigma_{Total}}$\\
		&   &    LTE  &  NLTE  & (dex) & $+$100\,K      & $+$0.2 dex     & $+$0.1 dex      & $+$0.2 $\rm{km}\,\rm{s^{-1}}$    &   (dex)  &  (dex) \\
		\hline\noalign{\smallskip}
		C (CH)   & \dots &  $-$0.20 		      &     \dots	                  &  \dots        &  \dots     &    \dots        &  \dots    &    \dots       &   \dots       &   \dots      \\  
 Na~I	  & 	4 &  	$+$0.26   &   $+$0.05          &    0.03 	& 0.04 	& $-$0.01 & 0.00  &	$-$0.07  &	0.08    &   0.11\\                        
Mg~I	  & 	3 & $+$0.31  &  $+$0.24          &    0.02 	& 0.02 	& $-$0.02 & 0.00  &	$-$0.06  &	0.07    &   0.09\\                           
Al~I	  & 	2 & $+$0.18 	  &   $+$0.08         &   0.01	& 0.07 	& 0.01 	& $-$0.01 & $-$0.03  &	0.08    &   0.09\\                             
Si~I	  & 	3 & $+$0.23    &  $+$0.22           &    0.05 	& $-$0.08 	& 0.03 	& 0.01  &	$-$0.04  &	0.09  &   0.14    \\                       
Ca~I	  & 	7 &  $-$0.02 &    $-$0.04       &    0.05 	& 0.07 	& 0.00 	& 0.02  &	$-$0.09  &	0.12    &   0.17\\                          
Sc~II   &	  4 & $+$0.12 	 &    \dots	                    &  	0.08 	& $-$0.01 	& 0.04 	& 0.02  &	$-$0.06  &	0.08  &   0.16\\              
Ti~II   &	  3 & $+$0.09 	 &    $+$0.08                    &  	0.04 	& $-$0.03 	& 0.04 	& 0.02  &	$-$0.07  &	0.09  &   0.13\\               
V I	    & 1 &   $+$0.10   & \dots & 	\dots     &  0.02     &   0.03  &  0.01  &  $-$0.06  &  0.07  &  0.07  \\                                      
V II    & 1 &	  $+$0.06  & \dots  & \dots     &  0.02     &   0.04  &  0.00  &  $-$0.07  &  0.08  &  0.08  \\                                       
Cr I    & 4 &	 $ - $0.02  &  $+$0.09 & 	0.02   &  $ - $0.01     &  0.04  &  0.02  &  $-$0.06  &  0.08  &  0.10\\                               
Mn I    & 3 &	 $+$0.10  &  $+$0.17  & 	0.01   &  0.03     &   0.01  &  0.02  &  $-$0.09  &  0.10  &  0.11\\                                      
Co I    & 1 &	 $+$0.00  &  $-$0.08 & 	\dots     &  0.01     &   0.03  &  0.01  &  $-$0.07  &  0.08  &  0.08  \\                                      
Ni~I	  & 	5 &  $-$0.10 &    \dots                    &    0.01 	& 0.01 	& 0.03 	& 0.00  &	$-$0.06  &	0.07    &   0.08\\                 
Cu~I	  & 	3 &  $-$0.07 &    $-$0.04      &    0.02 	& $-$0.01 	& 0.02 	& 0.01  &	$-$0.13  &	0.13  &   0.15\\                           
Zn I & 2 &	 $ - $0.19  &  \dots  & 	0.01   &  0.02     &   0.02  &  0.00  &  $-$0.10  &  0.10  &  0.11\\                                   
Sr~I	  & 	1 &  $-$0.09 &    \dots	                    &   \dots 	& 0.12 	& 0.00 	& 0.00  &	$-$0.12  &	0.17    &   0.17  \\                 
Y~II	  & 	3 &  $+$0.41   &    \dots	                    &    0.05 	& $-$0.02 	& 0.04 	& 0.02  &	$-$0.01  &	0.05  &   0.10\\ 
Zr~II   &	  3 &  $+$0.67 	  &   \dots	                    &  	0.02 	& $-$0.03 	& 0.04 	& 0.03  &	$-$0.02  &	0.06  &   0.08\\
Ba~II   &	  3 &  $+$0.42 	  &    $+$0.37           &  	0.08 	& 0.02 	& 0.03 	& 0.02  &	$-$0.12  &	0.13    &   0.21\\
La~II   &	  2 &  $+$0.57 	  &   \dots	                    &  	0.03 	& 0.03 	& 0.04 	& 0.03  &	$-$0.03  &	0.07    &   0.10\\
Ce~II   &	  3 &  $+$0.19 	  &   \dots	                    &  	0.10 	& 0.02 	& 0.02 	& 0.02  &	$-$0.09  &	0.10    &   0.20\\
Pr~II   &	  2 &  $+$0.65 	  &   \dots	                    &  	0.06 	& 0.03 	& 0.04 	& 0.02  &	$-$0.01  &	0.05    &   0.11\\
Nd~II   &	  3 &  $+$0.64 	  &   \dots	                    &  	0.03 	& 0.00 	& 0.04 	& 0.01  &	$-$0.10  &	0.11    &   0.14\\
Sm~II   &	  3 &  $+$1.10 	  &   \dots	                    &  	0.08 	& 0.02 	& 0.03 	& 0.03  &	$-$0.07  &	0.08    &   0.16\\
Eu~II   &	  2 &  $+$1.20 	  &    $+$1.32          &  	0.07 	& 0.00	& 0.04 	& 0.02  &	$-$0.03  &	0.05    &   0.12\\
Gd~II   &	  4 &  $+$1.14 	  &   \dots	                    &  	0.18 	& 0.01 	& 0.04 	& 0.02  &	$-$0.03  &	0.05    &   0.23\\
Dy~II   &	  1 &  $+$1.30 	  &   \dots	                    &  	\dots	& 0.02 	& 0.03 	& 0.00  &	$-$0.02  &	0.04    &   0.04  \\                   
Lu~II   &	  1 &  $+$0.20 	  &   \dots	                    &  	\dots	& $-$0.02 	& 0.03 	& 0.02  &	$-$0.01  &	0.04  &   0.04    \\         Hf~II   &   1  &  $+$1.10 &  \dots   &   \dots   &  $-$0.02  & 0.05    & 0.03    &  $-$0.02  & 0.06   &   0.06  \\  
Th~II   &  1    &  $+$1.00  &  \dots  &    \dots  &  0.02 &  0.02  & 0.03    &  $-$0.15  &  0.16    &   0.16  \\   
		\noalign{\smallskip}\hline 
	\end{tabular}
\end{table*}

\begin{figure*}[h]
	\centering
	\includegraphics[width=0.94\linewidth]{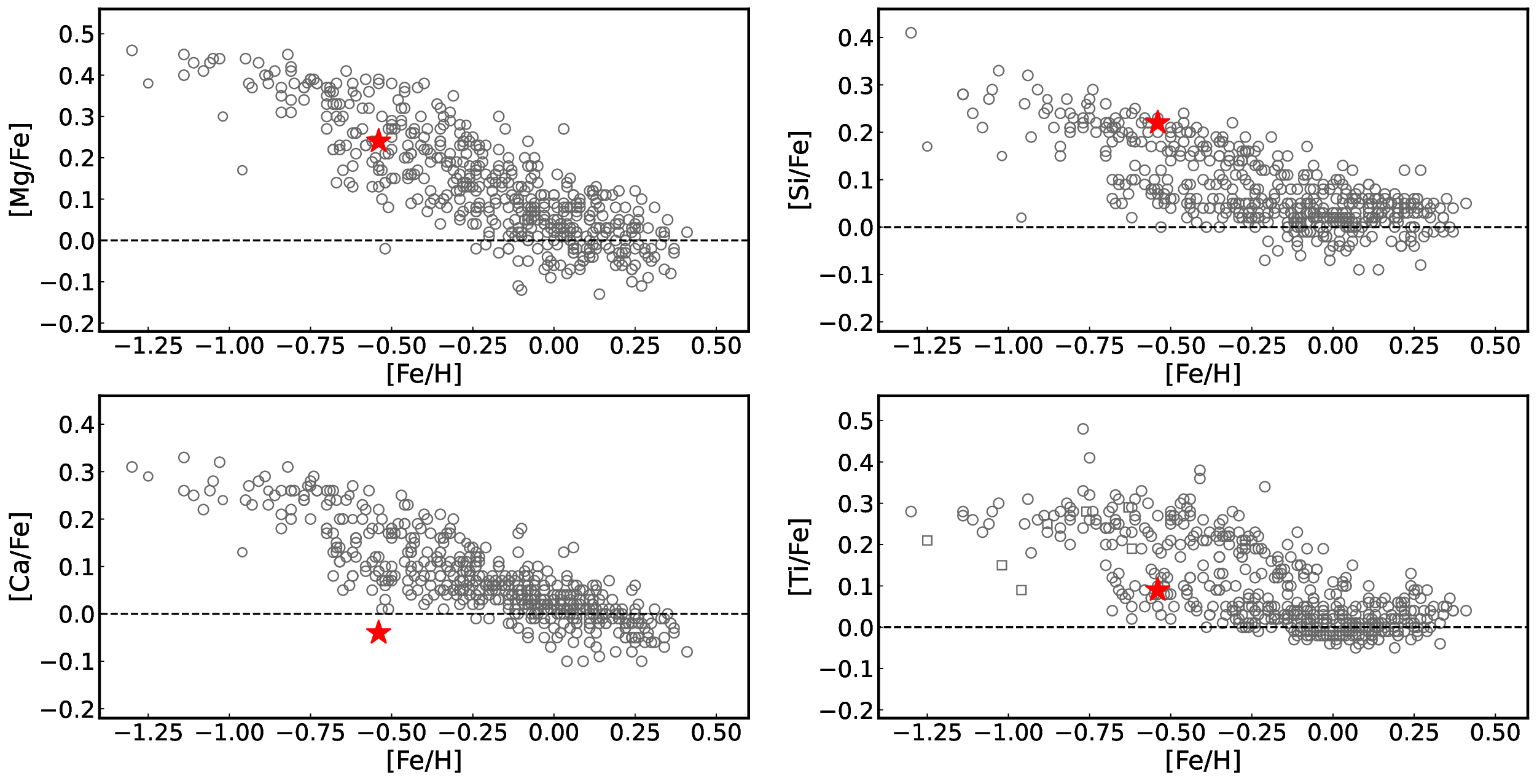}
	\includegraphics[width=0.94\linewidth]{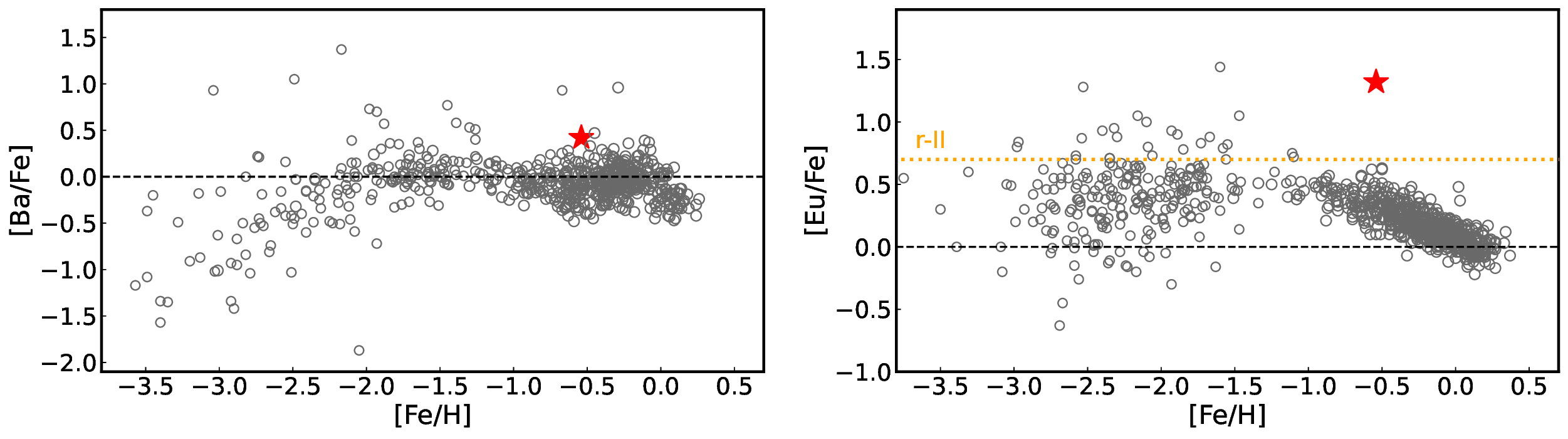}
	\caption{Comparison of abundance ratios in J\,0206$ + $4941 (red star) with MW field stars (gray circles). The abundance ratios of the MW stars come from \citet{2004AJ....128.1177V}, \citet{2014A&A...562A..71B}, \citet{2016A&A...586A..49B}, \citet{2019AA...631A.113F}, and \citet{2020ApJS..249...30H}. The black-dashed line is the Solar reference line. The orange-dashed line in the [Eu/Fe] vs. [Fe/H] panel indicates the $r$-II limit (+0.7) as defined by \citet{2020ApJS..249...30H}.
	\label{baeu}
    }

\end{figure*}

\begin{figure*}
	\centering
	\includegraphics[width=0.75\linewidth]{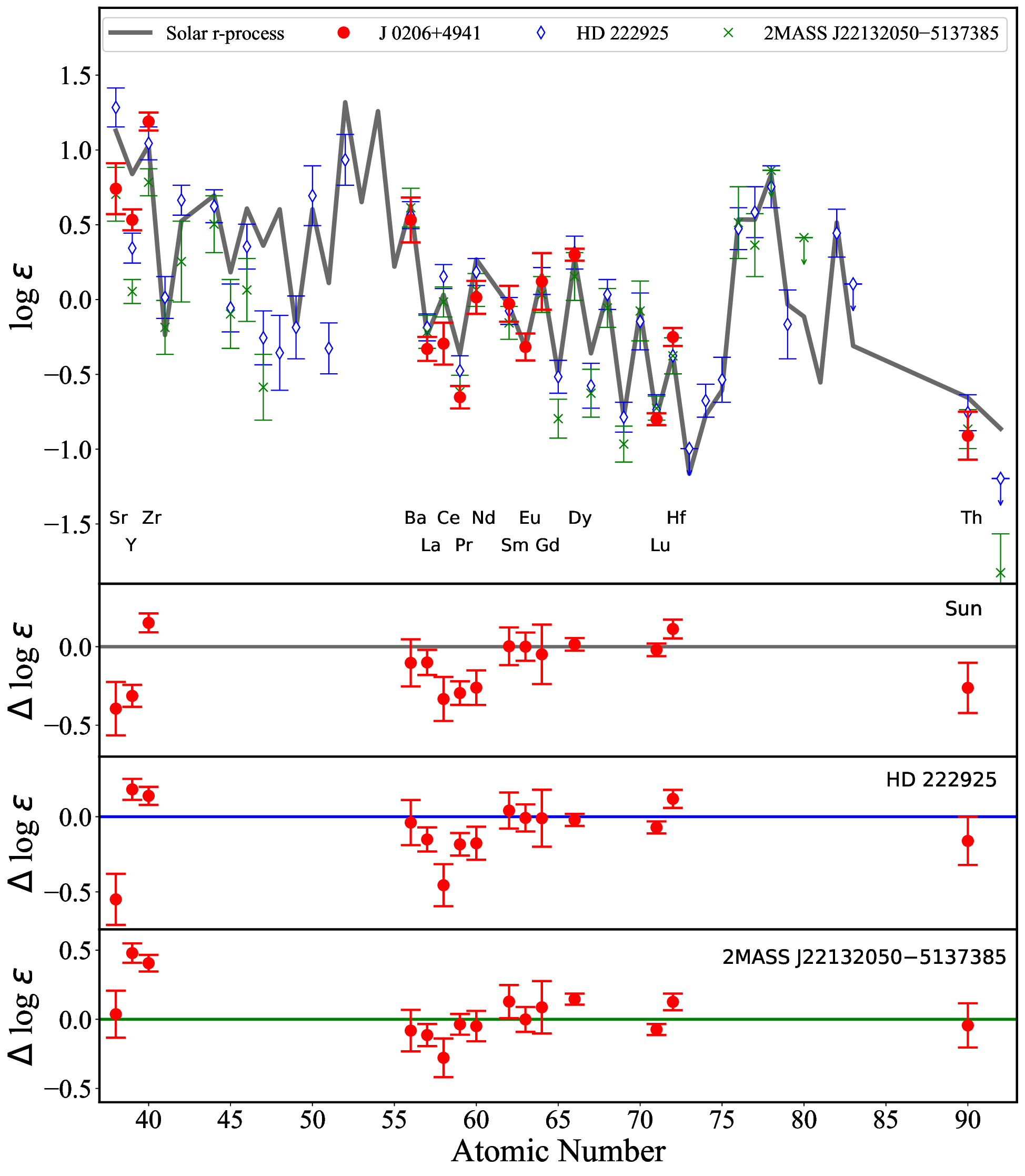}
	\caption{The heavy-element abundances of J\,0206$ + $4941 are shown in the top panel, with the Solar $r$-process pattern, abundances of HD 222925 \citep{2022ApJS..260...27R}, and 2MASS J22132050-5137385 \citep{2024arXiv240602691R} normalized to match the Eu abundance of J\,0206$ + $4941. Residuals between J\,0206$ + $4941 and the Solar $r$-process pattern, HD 222925, and 2MASS J22132050-5137385 are shown in the lower panels.}
    \vspace*{4mm} 
	\label{pattern}
\end{figure*}

\begin{figure}
	\centering
	\includegraphics[width=0.95\linewidth]{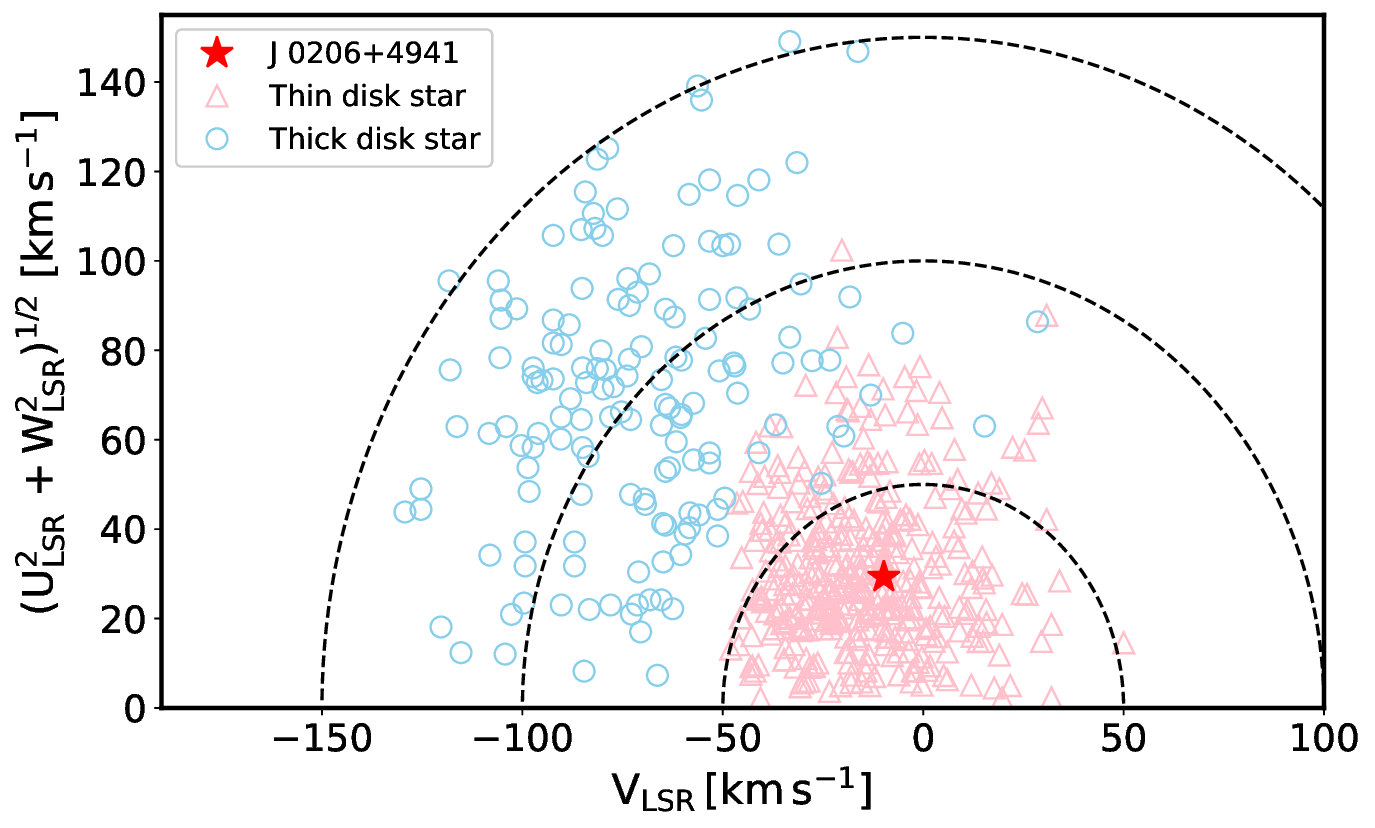}
	\caption{Toomre Diagram for J\,0206$ + $4941. The MW thin-disk stars (pink triangles) and thick-disk stars (blue circles) are from \citet{2003AA...410..527B} and \citet{2004AJ....128.1177V}, respectively.} 
	\label{uvw}
\end{figure}

\section{Discussion} \label{sec:disc}

\subsection{The Light and Fe-peak Elements}

The observed carbon abundance of J\,0206$ + $4941 is deficient, relative to the Solar value ([C/Fe] = $-$0.20). However, this star falls in the red giant region, therefore, its carbon has been depleted during the normal course of giant-branch evolution. Using the procedure described by \citet{2014ApJ...797...21P}, the carbon depletion for J\,0206$ + $4941 is $\sim$ 0.2\,dex, yielding a natal [C/Fe] $\approx$ 0.00, indicating that this star was likely never C-enhanced.
From inspection of Figure~\ref{baeu}, the abundances of Mg, Si, and Ti appear similar to those of the MW field stars. Although the [Ca/Fe] ratio is slightly lower, it is still close to the boundary of other field stars. The [Na/Fe] and [Al/Fe] ratios (not shown in the figure) are both slightly super-Solar. 

Among the Fe-group elements, [V/Fe], [Cr/Fe], and [Mn/Fe] are slightly over-abundant, while [Co/Fe], [Ni/Fe], [Cu/Fe], and [Zn/Fe] are slightly under-abundant. All of these abundance ratios are normal compared to stars of similar metallicity \citep[e.g.,][]{2015AA...577A...9B}.

\subsection{The $r$-process Enhancement and Pattern}

The lower two panels of Figure \ref{baeu} show the [Ba/Fe] and [Eu/Fe] ratios of J\,0206$ + $4941 compared to MW field stars. From inspection, this star exhibits a very strong enhancement of [Eu/Fe] ($+$1.32), and a slight enhancement of [Ba/Fe] ($+$0.37). The low [Ba/Eu] ratio ($-0.95$) indicates that its neutron-capture elements are mainly due to the $r$-process. 

In order to put J\,0206$ + $4941 in the context of other highly $r$-process-enhanced stars, we have listed the known RPE stars with [Eu/Fe] $\geq$ +1.0 in Table \ref{rpestars}, ordered by decreasing [Eu/H]. As shown in this table, most of these stars have low metallicity ([Fe/H] $\textless$ $ - $1.0) and are identified as members of the MW halo, while a few such stars are found in dwarf galaxies or globular clusters. J\,0206$ + $4941 is the first strongly $r$-process-enhanced star associated with the MW disk, and has a relatively high metallicity ([Fe/H] = $ - $0.54). The  [Eu/H] of J\,0206$ + $4941 is +0.78, which is the highest among known RPE stars. 

Many previous studies have found that the main $r$-process abundance patterns of metal-poor RPE stars are very close to that of the Solar $r$-process. This agreement is most noticeable for the elements from Ba to Hf, although there are some differences in light neutron-capture elements \citep[e.g.,][]{1994ApJ...431L..27S,2014AJ....147..136R, 2018ApJ...868..110S}, and star-to-star variations among the actinides. A comparison of the heavy-element abundances in J\,0206$+$4941 with the Solar $r$-process and two RPE ``template" stars (HD 222925; \citealt{2022ApJS..260...27R}, and 2MASS\,J22132050-5137385; \citealt{2024arXiv240602691R}) is shown in the top panel of Figure \ref{pattern}, where the Solar $r$-process, HD 222925, and 2MASS\,J22132050-5137385 $r$-process patterns are normalized to the Eu abundance of J\,0206$ + $4941. The residuals are presented in the bottom panels.

The abundance pattern of J\,0206$+$4941 is not perfectly consistent with the Solar $r$-process. The elements Sr, Y, Ce, Pr, and Nd are significantly lower compared to the Solar pattern, while only Sr and Ce are significantly lower compared to that of HD\,222925. When compared to 2MASS\,J22132050-5137385, the Ce abundance remains lower and the Y and Zr are higher. This difference suggests that the production of $r$-process elements may require multiple sites or multiple mechanisms, or that this star's abundances have a different contribution for these elements.

\subsection{Kinematics} 

We measure the heliocentric radial velocity ($\rm{{RV}_{helio}}$ = $ - $30.86 $\pm$ 0.67 km s$^ {-1}$) from the high-resolution GTC spectrum; it shows no obvious variation from the LAMOST DR7 \citep[$ - $30.94 $\pm$ 1.00 km\,s$^ {-1}$;][]{2022yCat.5156....0L}, and the \textit{Gaia} DR3 measurements \citep[$ - $30.00 $ \pm$ 0.40 km\,s$^{-1}$;][]{2023AA...674A...1G}. As a result, J\,0206$+$4941 is not likely to be in a binary system.

We calculate the local standard of rest (LSR) Galactic space velocities ($\rm{U_{LSR}}$, $\rm{V_{LSR}}$, $\rm{W_{LSR}}$)\, for J\,0206$+$4941, adopting the coordinates, proper motions, and parallax with zero-point offset corrected\footnote{The zero-point offset of the parallax is $-$0.014856 mas, as obtained using the package from \url{https://gitlab.com/icc-ub/public/gaiadr3\_zeropoint} \citep{2021AA...649A...4L}.} from the \textit{Gaia} DR3 catalog \citep{2023AA...674A...1G}, as well as the radial velocity derived from the GTC spectrum. The distance derived from the corrected parallax is $d=3.072\pm0.148\; \mathrm{kpc}$, which is consistent with the photogeometric distance from \citet[][$d=3.128\pm0.151\;\mathrm{kpc}$]{2021AJ....161..147B}. We use a right-handed Galactocentric Cartesian coordinate system and assume that the Sun is located at ($X_{\odot},Y_{\odot},Z_{\odot}$) = ($ - $8.2, 0.0, 0.0208) $ \mathrm{kpc}$ \citep{2017MNRAS.465...76M, 2019MNRAS.482.1417B}, the circular velocity at the Solar position is $v_0$ = 232.8 km\,s$^{-1}$ \citep{2017MNRAS.465...76M}, and the Solar peculiar motion is ($\rm{U_{\odot}}$, $\rm{V_{\odot}}$, $\rm{W_{\odot}}$) = (11.1, 12.24, 7.25) km s$^ {-1} $ \citep{2010MNRAS.403.1829S}. We adopt the MW potential from \citet{2017MNRAS.465...76M}, and employ the \texttt{AGAMA} routine \citep{2019MNRAS.482.1525V} to derive the orbital energy ($ E $), three-dimensional action ($ J = (J_r, J_{\phi}, J_z)$), pericentric radius (${r_{peri}}$), apocentric radius (${r_{apo}}$), maximum height from the Galactic midplane (${{Z}_{\mathrm{max}}}$), and eccentricity ($ecc = \frac{r_{apo}-r_{peri}}{r_{apo}+r_{peri}}$). 
We assume Gaussian distributions for its observables, and additionally account for the correlated observational uncertainties in proper motions and parallax. We run a Monte Carlo re-sampling procedure based on 100,000 draws to estimate the uncertainties of the dynamical parameters.

J\,0206$+$4941 exhibits typical thin-disk velocities (see Figure~\ref{uvw}) with a very low orbital eccentricity ($ecc = 0.054\pm0.004$) and limited orbital height (${{Z}_{\mathrm{max}} = 0.66\pm0.03}$ kpc). Many previously discovered highly $r$-process-enhanced stars have accretion origin \citep[e.g.,][]{2018AJ....156..179R, 2019NatAs...3..631X, 2019ApJ...874..148S}. However, J\,0206$+$4941 does not appear to be clustered in dynamical phase space with any known accreted structures \citep{2020ApJ...901...48N, 2021ApJ...912..157R, 2023ApJ...946...48H, 2023ApJ...943...23S}, thus it may well have been formed in-situ in the MW's thin disk. This star may have formed from an ISM enriched in $r$-process elements by a single binary neutron star merger and/or CCSNe, indicating that mixing may not be isotropic, even in the thin disk. We also note that \citet{Hirai2022} found, in their simulation of the origin of RPE stars in a MW-like galaxy, that all simulated $r-$II stars with [Fe/H] $> -1$ were formed in-situ, rather than via accretion from a parent dwarf satellite.

\subsection{Age Estimation} 

The presence of the radioactive element Th allows us to estimate the age of this star, albeit with a large error bar. The age can be estimated using the following relation \citep{2001Natur.409..691C}:
\[ t = 46.67\,{\rm Gyr\,[\,log\,\epsilon\,(Th/Eu)_0 - log\,\epsilon\,(Th/Eu)_{obs}],} \]
where $\rm log\epsilon(Th/Eu)_0 $ is the initial production ratio (PR) corresponding to the element formation at $t =$ 0, and $\rm log\epsilon(Th/Eu)_{obs} $ is the observed ratio after the radioactive element Th has decayed within a time $ t $. We adopted an initial PR from \citet{2002ApJ...579..626S}, and the calculated age is 12.32\,Gyr. The uncertainty is estimated only from the propagation of the abundance-measurement uncertainty. Only one Th line can be measured, and the Th abundance has a systematic error of 0.16\,dex, which leads to a large age uncertainty of 9.08\,Gyr. 

\vskip 1cm
\section{Conclusions}\label{sec:conc}

We describe the discovery of an extremely $r$-process-enhanced thin-disk star, J\,0206$+$4941. It was originally selected from the LAMOST medium-resolution survey as a candidate RPE star, based on its unusually strong lines of Eu. A high-S/N, moderately high-resolution ($R \sim 25,000$) follow-up observation was obtained using HORuS on the GTC. We derive the abundances of 30 elements, and determine the star's kinematics. Our main findings can be summarized as follows:

\begin{enumerate}
\item J\,0206$+$4941 is the most metal-rich ([Fe/H] = $ - $0.54) highly $r$-process-enhanced star found to date, and exhibits the highest known abundance ratio of the $r$-process element Eu relative to H ([Eu/H] = +0.78). 

\item The abundances of the light elements in this star with $Z \leq$ 30 are commensurate with other stars of similar metallicity and evolutionary phase, with a strong enhancement of Eu, ([Eu/Fe] = +1.32) and modest enhancement of Ba ([Ba/Fe] = +0.37). The observed [Ba/Eu] ratio ([Ba/Eu] = $ - $0.95) indicates that there is relatively little contribution from the $s$-process to the enrichment of the neutron-capture elements. 
 
\item The abundance pattern of the elements with $Z \geq$ 30 is not perfectly consistent with the scaled Solar $r$-process pattern, as demonstrated by the lower abundances of Ce, Pr, and Nd. This difference provides some support for multiple origins of the $r$-process. 

\item J\,0206$+$4941 does not exhibit significant radial-velocity variations, indicating it is not likely to be in a binary system.

\item The Galactic space velocities and orbital parameters show that J\,0206$+$4941 is restricted to the thin disk of the MW, and apparently was not accreted from an external dwarf galaxy. 

\item The presence of the radioactive element Th allows us to estimate the age of J\,0206$+$4941 as 12.32 $\pm$ 9.08 Gyr.

\end{enumerate}

Clearly, a higher-resolution ($R > 50,000$) spectrum of this star with better S/N in the blue region would prove illuminating, and enable the measurement of numerous other $r$-process elemental abundances, possibly including uranium, enabling derivation of a more accurate age estimate.
In addition to J\,0206$+$4941, we have also found a few other candidate RPE stars with [Fe/H] $> -$1.0 from the LAMOST-MRS spectra. These stars are excellent candidates for high-resolution follow-up spectroscopic observations with HORuS and other instruments in the near future.
 
\vskip 0.4cm 
\noindent We thank the anonymous referee for valuable suggestions that have improved the manuscript. This research is supported by the National Natural Science Foundation of China under Grant Nos. 12090040, 12090044, 12373036, 12022304, 11973052, the National Key R$\rm\& $D Program of China No.2019YFA0405502, the Scientific Instrument Developing Project of the Chinese Academy of Sciences, Grant No.\,ZDKYYQ20220009, and the International Partnership Program of the Chinese Academy of Sciences, Grant No.\,178GJHZ2022047GC. H.-L.Y. acknowledges the support from the Youth Innovation Promotion Association, Chinese Academy of Sciences. CAP is thankful for financial support from the Spanish Ministry MICINN projects AYA2017-86389-P and PID2020-117493GB-I00. T.C.B. acknowledges support from grant PHY 14-30152; Physics Frontier Center/JINA Center for the Evolution of the Elements (JINA-CEE), and from OISE-1927130: The International Research  Network for Nuclear Astrophysics (IReNA), awarded by the US National Science Foundation.
 
This paper is based on observations made with the Gran Telescopio Canarias (GTC), installed at the Spanish Observatorio del Roque de los Muchachos of	the Instituto de Astrofísica de Canarias, on the island of La Palma. The Guoshoujing Telescope (the Large Sky Area Multi-Object Fiber Spectroscopic Telescope LAMOST) is a National Major Scientific Project built by the Chinese Academy of Sciences. Funding for the project has been provided by the National Development and Reform Commission. LAMOST is operated and managed by the National Astronomical Observatories, Chinese Academy of Sciences.

\bibliography{xie_gtc}{}
\bibliographystyle{aasjournal}

\setcounter{table}{0}
\renewcommand{\thetable}{A\arabic{table}}

\begin{appendix}

Here we assemble a list of highly r-process-enhanced stars, chosen to have [Eu/Fe] $\geq + 1.0$, sorted by decreasing [Eu/H]. Note that the abundances listed in the table are LTE results. The [C/Fe] listed is the ``as measured" quantity, not including corrections for evolutionary state.  The column labeled LOC indicates the likely stellar population with which each star is associated.

 \startlongtable
\begin{deluxetable*}{cccccrrrrrcc} 	
\tablecaption{ Highly r-process-enhanced Stars ([Eu/Fe] $\geq +1.0$ and [Ba/Eu] $\textless$ 0.0). } 
        \label{rpestars}
	\tablehead{\colhead{Stellar ID} & \colhead{$ {T_{\rm eff}} $} & \colhead{$\rm log\,g $} & \colhead{$\xi_{\rm t}$} & \colhead{$ \rm[Fe/H] $} & \colhead{$ \rm[Eu/H] $} & \colhead{$ \rm[Eu/Fe] $}   & \colhead{$ \rm[Ba/Eu] $} & \colhead{$ \rm[C/Fe] $} & \colhead{$ \rm[Mg/Fe] $} & \colhead{LOC} & \colhead{REF} \\ 
		\colhead{} & \colhead{(K)} & \colhead{(cgs)} & \colhead{($ \rm km\,{s^{\rm -1}} $)} & \colhead{} & \colhead{} & \colhead{} & \colhead{} & \colhead{} & \colhead{} & \colhead{} & \colhead{} 
	}  	
	\startdata
 LAMOST J020632.21+494127.9 & 4130 & 1.52 & 1.80 & $-$0.54 & +0.78 & +1.32 & $-$0.95 & $-$0.20 & +0.31 & Disk  &  this work\\
$\rm{[LDH2014]}$Fnx-mem0556	&	4176	&	0.70 	&		&	$-$0.79 	&	+0.46 	&	+1.25 	&     $-$0.88 	&	\dots	&	$-$0.15 	&	Fnx\tablenotemark{ a}	&	REI21	\\                                               
$\rm{[LDH2014]}$Fnx-mem0595	&	4223	&	0.66 	&		&	$-$0.92 	&	+0.46 	&	+1.38 	&     $-$0.85 	&	\dots		&	$-$0.17 	&	Fnx\tablenotemark{ a}	&	REI21	\\
	BPS CS 31070-073	&	6190	&	3.86 	&	1.50 	&	$-$2.55 	&	+0.28 	&	+2.83 	&              $-$0.41 	&	+1.34 	&	+0.64 	&	Halo	&	ALL12	\\                                  
	2MASS J22132050-5137385	&	5509	&	2.28 	&	1.25 	&	$-$2.20 	&	+0.25 	&	+2.45 	&      $-$0.73 	&	+0.38 	&	+0.49 	&	Halo	&	ROE24	\\                                  
	$\rm{[LDH2014]}$Fnx-mem0546	&	4367	&	0.82 	&		&	$-$1.33 	&	+0.12 	&	+1.45 	&                      $-$1.05 	&	\dots		&	$-$0.11 	&	Fnx\tablenotemark{ a}	&	REI21	\\                                  
	BPS CS 29526-110	&	6650	&	3.79 	&	2.10 	&	$-$2.19 	&	+0.09 	&	+2.28 	&              $-$0.01 	&	+2.38 	&	+0.22 	&	Halo	&	ALL12	\\                  UMi COS 92 &	4325	& 0.30 &	1.95	& $-$1.45 & 	+0.04 &	+1.49  &	$-$0.72 &	\dots	 &	+0.02 &	UMi\tablenotemark{ b}	& SHE01		\\ %
	HD 222925	&	5636	&	2.54 	&	2.20 	&	$-$1.47 	&	$-$0.14 	&	+1.33 	&                    $-$0.78 	&	$-$0.20 	&	+0.41 	&	Halo	&	ROE18	\\                                
	2MASS J07202253-3358518	&	5040	&	2.22 	&	1.78 	&	$-$1.60 	&	$-$0.16 	&	+1.44 	&    $-$0.76 	&	+0.11 	&	\dots		&	Halo	&	HOL20	\\                                      
	LAMOST J112456.61+453531.3	&	5180	&	2.70 	&	1.50 	&	$-$1.27 	&	$-$0.17 	&	+1.10 	&  $-$0.86 	&	$-$0.41 	&	$-$0.31 	&	Halo	&	XIN19	\\                              
	RAVE J040618.2-030525	&	5260	&	2.75 	&	1.80 	&	$-$1.34 	&	$-$0.17 	&	+1.17 	&        $-$0.19 	&	+0.54 	&	\dots		&	Halo	&	RAS20	\\                                      
	DES J033457.57-540531.4	&	5328	&	2.85 	&	1.50 	&	$-$2.08 	&	$-$0.31 	&	+1.77 	&    $-$0.41 	&		&		&	Ret II\tablenotemark{ c}	&	JI16	\\                                        
	2MASS J07103110-7121522	&	5167	&	2.94 	&	1.75 	&	$-$1.47 	&	$-$0.42 	&	+1.05 	&    $-$0.72 	&	0.00 	&	\dots		&	Halo	&	HOL20	\\                                      
	SMSS J175046.30-425506.9	&	4752	&	1.55 	&	2.05 	&	$-$2.17 	&	$-$0.42 	&	+1.75 	&    $-$0.90 	&	$-$0.12 	&	+0.24 	&	Halo	&	JAC15	\\                                
	RAVEJ153830.9-180424	&	4995	&	2.00 	&	1.85 	&	$-$2.02 	&	$-$0.50 	&	+1.52 	&        $-$0.80 	&	+0.12 	&		&	Halo	&	RAS20	\\                                      
	2MASS J18024226-4404426	&	4701	&	1.60 	&	2.17 	&	$-$1.55 	&	$-$0.50 	&	+1.05 	&    $-$0.10 	&	+0.35 	&	\dots		&	Halo	&	HAN18	\\                                      
	2MASS J15213995-3538094	&	5850	&	2.10 	&	2.65 	&	$-$2.80 	&	$-$0.57 	&	+2.23 	&    $-$0.88 	&	+0.56 	&	+0.37 	&	Halo	&	CAI20	\\                                  
	SMSS J183128.71-341018.4	&	4940	&	2.15 	&	2.00 	&	$-$1.83 	&	$-$0.58 	&	+1.25 	&    $-$0.72 	&	+0.01 	&	+0.22 	&	Bulge	&	HOW16	\\                                  
	2MASS J19570805+5529491&	5070	&	1.30 	&	2.18 	&	$-$1.72 	&	$-$0.61 	&	+1.11 	&$-$1.26 	&	\dots		&	+0.31 	&	Halo	&	HAW18	\\                                      
	  2MASS J07114252-3432368   	&	4767	&	1.33 	&	1.74 	&	$-$1.96 	&	$-$0.66 	&	+1.30 	&            $-$0.80 	&	+0.47 	&	+0.31 	&	Halo	&	SAK18	\\                                  
	SMSS J183225.29-334938.4	&	5293	&	2.35 	&	1.80 	&	$-$1.74 	&	$-$0.66 	& +1.08 	&    $-$0.58 	&	$-$0.22 	&	+0.42 	&	Bulge	&	HOW16	\\                                                                        
	2MASS J05241392-0336543	&	4430	&	2.25 	&	3.00 	&	$-$2.20 	&	$-$0.70 	&	+1.50 	&    $-$1.03 	&	$-$0.40 	&	$-$0.14 	&	Halo	&	EZZ20	\\                              
	2MASS J03073894-0502491	&	4360	&	1.75 	&	2.90 	&	$-$1.98 	&	$-$0.71 	&	+1.27 	&    $-$1.13 	&	$-$0.57 	&	+0.09 	&	Halo	&	EZZ20	\\                                
	2MASS J13052137-1137220	&	4600	&	1.80 	&	2.20 	&	$-$1.82 	&	$-$0.73 	&	+1.09 	&    $-$0.89 	&	$-$0.30 	&	$-$0.59 	&	Halo	&	EZZ20	\\                              
	2MASS J20093393-3410273	&	4690	&	1.64 	&	1.70 	&	$-$2.10 	&	$-$0.78 	&	+1.32 	&    $-$1.09 	&	$-$0.86 	&	\dots		&	Halo	&	HAN18	\\                                                                                                     
	2MASS J00101758-1735387	&	5200	&	3.00 	&	1.70 	&	$-$2.19 	&	$-$0.84 	&	+1.35 	&    $-$0.92 	&	0.00 	&	$-$0.56 	&	Halo	&	EZZ20	\\                                
	2MASS J12170829+0415146	&	4540	&	1.65 	&	2.90 	&	$-$2.00 	&	$-$0.90 	&	+1.10 	&    $-$0.95 	&	$-$0.40 	&	$-$0.18 	&	Halo	&	EZZ20	\\                              
	2MASS J23362202-5607498	&	4630	&	1.28 	&	2.15 	&	$-$2.06 	&	$-$0.92 	&	+1.14 	&    $-$0.90 	&	$-$0.05 	&	\dots		&	Halo	&	HAN18	\\                                    
	2MASS J00512646-1053170	&	6440	&	4.02 	&	1.59 	&	$-$2.34 	&	$-$0.97 	&	+1.37 	&    $-$0.72 	&	+0.62 	&	+0.45 	&	Halo	&	SHA24	\\                                  
	RAVE J012931.1-160046	&	4959	&	1.70 	&	2.65 	&	$-$2.77 	&	$-$0.97 	&	+1.80 	&        $-$0.60 	&	+0.44 	&	+0.47 	&	Halo	&	RAS20	\\                                                     
	2MASS J14534137+0040467	&	4550	&	1.55 	&	2.30 	&	$-$2.82 	&	$-$1.02 	&	+1.80 	&    $-$0.75 	&	$-$0.41 	&	+0.92 	&	Halo	&	EZZ20	\\                                
	DES J033537.06-540401.2	&	5170	&	2.45 	&	1.55 	&	$-$2.73 	&	$-$1.03 	&	+1.70 	&    $-$0.30 	&	\dots		&	\dots		&	Ret II\tablenotemark{ c}	&	JI16	\\                                        
	DES J033447.93-540525.0	&	4900	&	1.70 	&	1.90 	&	$-$2.91 	&	$-$1.04 	&	+1.87 	&    $-$0.79 	&	\dots		&	\dots		&	Ret II\tablenotemark{ c}	&	JI16	\\                                                                
	HE 0048-1109	&	6265	&	3.80 	&	1.45 	&	$-$2.35 	&	$-$1.06 	&	+1.29 	&                $-$0.67 	&	+0.61 	&	+0.30 	&	Halo	&	GUL21	\\              
	CS 31082-001   	&	4827	&	1.65 	&	1.70 	&	$-$2.79 	&	$-$1.07 	&	+1.72 	&            $-$0.50 	&	+0.04 	&	+0.46 	&	Halo	&	SAK18	\\                                  
	2MASS J03270229+0132322	&	5100	&	2.95 	&	1.80 	&	$-$2.17 	&	$-$1.10 	&	+1.07 	&    $-$0.57 	&	+0.35 	&	+0.07 	&	Halo	&	EZZ20	\\                                  
	HE 0007-1752	&	5215	&	2.65 	&	1.45 	&	$-$2.36 	&	$-$1.11 	&	+1.25 	&                $-$0.65 	&	$-$0.29 	&	$-$0.33 	&	Halo 	&	GUL21	\\
	2MASS J03422816-6500355	&	4976	&	2.29 	&	2.05 	&	$-$2.16 	&	$-$1.11 	&	+1.05 	&    $-$0.65 	&	+0.05 	&	\dots		&	Halo	&	HOL20	\\                                      
	2MASS J15260106-0911388	&	4499	&	0.76 	&	2.41 	&	$-$2.83 	&	$-$1.13 	&	+1.70 	&    $-$1.01 	&	$-$0.82 	&	\dots		&	Halo	&	HAN18	\\                                    
	DES J033556.28-540316.3	&	5305	&	2.95 	&	1.65 	&	$-$3.54 	&	$-$1.14 	&	$\textless$+2.40	&        $-$2.30 	&	\dots		&	\dots		&	Ret II\tablenotemark{ c}	&	JI16	\\                                        
	2MASS J21091825-1310062	&	4855	&	1.42 	&	2.10 	&	$-$2.40 	&	$-$1.15 	&	+1.25 	&            $-$1.13 	&	$-$0.28 	&	\dots		&	Halo	&	HAN18	\\                                                                     
	J0246-1518     	&	4948	&	1.93 	&	1.56 	&	$-$2.45 	&	$-$1.16 	&	+1.29 	&                    $-$0.64 	&	$-$0.07 	&	+0.13 	&	Halo	&	SAK18	\\                                
	2MASS J14325334-4125494	&	5020	&	2.39 	&	1.80 	&	$-$2.79 	&	$-$1.18 	&	+1.61 	&            $-$0.90 	&	$-$0.14 	&	\dots		&	Halo	&	HAN18	\\                                    
	CS 29497-004	&	5013	&	2.23 	&	1.62 	&	$-$2.81 	&	$-$1.19 	&	+1.62 	&                        $-$0.41 	&	+0.22 	&	+0.31 	&	Halo	&	BAR05	\\                                  
	HE 0324-0122	&	5145	&	2.45 	&	1.65 	&	$-$2.41 	&	$-$1.20 	&	+1.21 	&                        $-$0.75 	&	+0.37 	&	+0.36 	&	Halo	&	GUL21	\\              
	2MASS J22182082-3827554	&	5000	&	2.70 	&	1.60 	&	$-$2.32 	&	$-$1.22 	&	+1.10 	&            $-$0.60 	&	+0.30 	&	+0.18 	&	Halo	&	EZZ20	\\                                  
	DES J033607.74-540235.5	&	4833	&	1.55 	&	2.15 	&	$-$2.97 	&	$-$1.23 	&	+1.74 	&           $-$0.83 	&	 \dots		&	\dots		&	Ret II\tablenotemark{ c}	&	JI16	\\                                        
	2MASS J05383296-5904280	&	5824	&	2.03 	&	2.84 	&	$-$2.53 	&	$-$1.25 	&	+1.28 	&            $-$0.52 	&	$\textless$+0.80	&	\dots		&	Halo	&	HOL20	\\                          
	2MASS J02462013-1518419	&	4879	&	1.80 	&	2.35 	&	$-$2.71 	&	$-$1.26 	&	+1.45 	&            $-$0.85 	&	+0.04 	&	\dots		&	Halo	&	HAN18	\\                                      
	HE 1523-0901   	&	4290	&	0.20 	&	2.13 	&	$-$3.09 	&	$-$1.27 	&	+1.82 	&                    $-$2.37 	&	+0.39 	&	+0.47 	&	Halo	&	SAK18	\\                                  
	2MASS J19161821-5544454	&	4450	&	0.65 	&	2.50 	&	$-$2.35 	&	$-$1.27 	&	+1.08 	&            $-$1.20 	&	$-$0.80 	&	\dots		&	Halo	&	HAN18	\\                                    
	RAVE J203843.2–002333	&	4630	&	1.20 	&	2.15 	&	$-$2.91 	&	$-$1.27 	&	+1.64 	&              $-$0.81 	&	$-$0.44 	&	+0.36 	&	Halo	&	PLA17	\\                                
	DES J033523.85-540407.5	&	4608	&	1.00 	&	2.40 	&	$-$3.01 	&	$-$1.33 	&	+1.68 	&            $-$0.89 	&	 \dots		&	\dots		&	Ret II\tablenotemark{ c}	&	JI16	\\                                        
	2MASS J17225742-7123000	&	5080	&	2.67 	&	0.70 	&	$-$2.42 	&	$-$1.35 	&	+1.07 	&            $-$0.37 	&	$-$0.33 	&	\dots		&	Halo	&	HAN18	\\                                    
	M15 31914	&	4338	&	0.58 	&	2.20 	&	$-$2.41 	&	$-$1.35 	&	+1.06 	&                              $-$0.70 	&	\dots		&	\dots		&	M15\tablenotemark{ d}	&	GAR24	\\                                        
	HE 1226-1149	&	5120	&	2.35 	&	1.50 	&	$-$2.91 	&	$-$1.36 	&	+1.55 	&                        $-$0.65 	&	+0.42 	&	+0.32 	&	Halo	&	COH13	\\                                  
	2MASS J02165716-7547064	&	4543	&	0.71 	&	2.44 	&	$-$2.50 	&	$-$1.38 	&	+1.12 	&             $-$0.87 	&	$-$0.33 	&		&	Halo	&	HAN18	\\                                                                   
	2MASS J21064294-6828266	&	5186	&	2.70 	&	2.90 	&	$-$2.76 	&	$-$1.44 	&	+1.32 	&             $-$0.80 	&	+0.53 	&	\dots		&	Halo	&	HAN18	\\                                      
	SMSS J175738.37-454823.5	&	4617	&	1.20 	&	2.30 	&	$-$2.46 	&	$-$1.44 	&	+1.02 	&             $-$3.01 	&	$-$0.43 	&	+0.39 	&	Halo	&	JAC15	\\                                
	SDSS J235718.91-005247.8	&	5000	&	4.80 	&	0.00 	&	$-$3.36 	&	$-$1.44 	&	+1.92 	&             $-$0.80 	&	+0.43 	&	+0.19 	&	Halo	&	AOK10	\\                                  
	BPS CS 22892-052	&	4800	&	1.50 	&	1.95 	&	$-$3.10 	&	$-$1.46 	&	+1.64 	&                     $-$0.65 	&	+0.88 	&	+0.30 	&	Halo	&	SNE03	\\                                  
	SMSS J155430.57-263904.8	&	4783	&	1.20 	&	2.20 	&	$-$2.61 	&	$-$1.47 	&	+1.14 	&             $-$0.34 	&	$-$0.26 	&	+0.33 	&	Halo	&	JAC15	\\                                                                 
	HE 2224+0143	&	5198	&	2.66 	&	1.67 	&	$-$2.58 	&	$-$1.53 	&	+1.05 	&                         $-$0.46 	&	+0.35 	&	+0.32 	&	Halo	&	BAR05	\\                                  
	HE 0430-4901	&	5296	&	3.12 	&	1.32 	&	$-$2.72 	&	$-$1.56 	&	+1.16 	&                         $-$0.66 	&	+0.09 	&	+0.17 	&	Halo	&	BAR05	\\                                  
	HE 1219-0312	&	5060	&	2.30 	&	1.60 	&	$-$2.96 	&	$-$1.58 	&	+1.38 	&                         $-$0.73 	&	+0.03 	&	+0.41 	&	Halo	&	HAY09	\\                                  
	BPS CS 22958-052	&	6090	&	3.75 	&	1.95 	&	$-$2.62 	&	$-$1.62 	&	+1.00 	&                     $-$1.00 	&	+0.32 	&	\dots		&	Halo	&	ROE14	\\                                      
	BPS CS 31078-018	&	5257	&	2.75 	&	1.50 	&	$-$2.85 	&	$-$1.62 	&	+1.23 	&                     $-$0.51 	&	+0.37 	&	+0.42 	&	Halo	&	LAI08	\\                                  
	HE 1127-1143	&	5224	&	2.64 	&	1.59 	&	$-$2.73 	&	$-$1.65 	&	+1.08 	&                         $-$0.45 	&	+0.54 	&	+0.22 	&	Halo	&	BAR05	\\                                  
	SMSS J062609.83-590503.2	&	4960	&	1.75 	&	1.75 	&	$-$2.77 	&	$-$1.71 	&	+1.06 	&             $-$0.22 	&	+0.27 	&	+0.24 	&	Halo	&	JAC15	\\                                  
	HE 2327-5642	&	5048	&	2.22 	&	1.69 	&	$-$2.95 	&	$-$1.73 	&	+1.22 	&                         $-$0.56 	&	+0.43 	&	+0.14 	&	Halo	&	BAR05	\\                                  
	CS 29491-069	&	5103	&	2.45 	&	1.54 	&	$-$2.81 	&	$-$1.75 	&	+1.06 	&                         $-$0.72 	&	+0.18 	&	+0.28 	&	Halo	&	BAR05	\\                                  
	BPS CS 22945-017	&	6080	&	3.70 	&	1.25 	&	$-$2.89 	&	$-$1.76 	&	+1.13 	&                     $-$0.64 	&	+1.78 	&	\dots		&	Halo	&	ROE14	\\                                      
	DES J033531.13-540148.2	&	4925	&	1.90 	&	1.80 	&	$-$3.34 	&	$-$1.84 	&	$\textless$+1.50	&   $-$2.30 	&	\dots		&	\dots		&	Ret II\tablenotemark{ c}	&	JI16	\\                                        
	BPS CS 22945-058	&	5990	&	3.65 	&	1.55 	&	$-$2.98 	&	$-$1.84 	&	+1.14 	&                     $-$0.85 	&	+0.68 	&	\dots		&	Halo	&	ROE14	\\                                      
	HE 0432-0923	&	5131	&	2.64 	&	1.54 	&	$-$3.19 	&	$-$1.94 	&	+1.25 	&                         $-$0.53 	&	+0.24 	&	+0.34 	&	Halo	&	BAR05	\\                                  
	BPS BS 16929-005	&	5250	&	3.10 	&	2.30 	&	$-$3.15 	&	$-$2.14 	&	+1.01 	&                     $-$0.74 	&	+1.50 	&	+0.49 	&	Halo	&	ALL12	\\                                  
	SPLUS J142445.34-254247.1	&	4762	&	1.58 	&	1.60 	&	$-$3.82 	&	$-$2.20 	&	+1.62 	&           $-$0.37 	&	$-$0.21 	&	+0.54 	&	Halo	&	PLA23	\\                                
	LAMOST J110901.22+075441.8	&	4440	&	0.70 	&	1.98 	&	$-$3.41 	&	$-$2.25 	&	+1.16 	&           $-$0.85 	&	$-$0.57 	&	+0.41 	&	Halo	&	LIH15	\\
 	BPS CS 22885-096	&	5050	&	2.00 	&	1.80 	&	$-$3.60 	&	$-$2.60 	&	+1.00 	&   $-$1.74 	&	  \dots  	&	+0.26 	&	Halo	&	RYA96	\\          
	SMSS J024858.41-684306.4	&	4977	&	1.60 	&	1.80 	&	$-$3.71 	&	$-$2.71 	&	+1.00 	&     $-$0.41 	&	+0.66 	&	+0.57 	&	Halo	&	JAC15	\\    
	\enddata

    	\tablenotetext{a} {The Fornax dwarf galaxy.}
        \tablenotetext {b} {The Ursa Minor dwarf galaxy.} 
    	\tablenotetext{c} {The Reticulum II dwarf galaxy.}
        \tablenotetext{d} {The M15 globular cluster.}
    
	\tablecomments{RYA96: \citet{Ryan_1996}; SHE01: \citet{2001ApJ...548..592S}; SNE03: \citet{2003ApJ...591..936S}; BAR05: \citet{2005AA...439..129B}; LAI08: \citet{2008ApJ...681.1524L}; HAY09: \citet{2009AA...504..511H}; AOK10:  \citet{2010ApJ...723L.201A}; ALL12: \citet{2012AA...548A..34A}; COH13: \citet{2013ApJ...778...56C}; ROE14: \citet{2014MNRAS.445.2946R}; JAC15: \citet{2015ApJ...807..171J}; LIH15: \citet{2015RAA....15.1264L}; HOW16: \citet{2016MNRAS.460..884H}; JI16: \cite{2016Natur.531..610J}; PLA17: \citet{Placco_2017}; HAN18: \citet{2018ApJ...858...92H}; HAW18: \citet{2018MNRAS.481.1028H}; ROE18: \citet{2018ApJ...865..129R}; SAK18: \citet{2018ApJ...868..110S}; XIN19: \citet{2019NatAs...3..631X}; CAI20: \citet{Cain_2020}; EZZ20: \citet{2020ApJ...898..150E}; HOL20: \citet{2020ApJS..249...30H}; RAS20: \citet{2020ApJ...905...20R}; GUL21: \citet{2021ApJ...912...52G}; REI21: \citet{2021ApJ...912..157R}; PLA23: \citet{Placco_2023}; GAR24: \citet{2024ApJ...967..101C}; ROE24: \citet{2024arXiv240602691R}; SHA24: \citet{2024MNRAS.529.1917S}. 
 }

\end{deluxetable*}

\end{appendix}

\end{document}